\documentclass[useAMS,usenatbib,usegraphicx]{mn2e}			%include twocolumn before 10pt to split into two columns
\usepackage{times}
\usepackage{natbib} 
\usepackage[dvips]{graphicx,color}
\usepackage{amsmath}
\usepackage{amssymb}
\usepackage[normalem]{ulem}

\title[Spectroscopic bulge--disc decomposition]{Spectroscopic bulge--disc decomposition: \\a new method to study the evolution of lenticular galaxies}
\author[E.J. Johnston et al.]{E.J.~Johnston,$^1$\thanks{Email: ppxej@nottingham.ac.uk} A.~Arag\'on-Salamanca,$^1$ M.R.~Merrifield$^1$ and A.G.~Bedregal$^{2,3}$\\
$^1$School of Physics and Astronomy, University of Nottingham, University Park, Nottingham, NG7 2RD, UK\\
$^2$Departamento de Astrofisica, Facultad de Ciencias Fisicas, Universidad Computense de Madrid, 28040 Madrid, Spain\\
$^3$Institute for Astrophysics, University of Minnesota, 116 Church Street SE, Minneapolis, MN 55455 USA\\}

\begin{document}

\maketitle
\begin{abstract}
  A new method for spectroscopic bulge--disc decomposition is
  presented, in which the spatial light profile in a two-dimensional
  spectrum is decomposed wavelength-by-wavelength into bulge and disc
  components, allowing separate one-dimensional spectra for each
  component to be constructed. This method has been applied to
  observations of a sample of nine lenticular galaxies in the Fornax
  Cluster in order to obtain clean high-quality spectra of their
  individual bulge and disc components.  So far this decomposition 
  has only been fully successful when applied to galaxies with clean light
  profiles, without contamination from dust lanes etc. This has consequently 
  limited the number of galaxies that could be separated into bulge and disc 
  components. Lick index stellar 
  population analysis of the component spectra reveals that in those 
  galaxies where the bulge and disc could be distinguished, the bulges 
  have systematically  higher metallicities and younger stellar 
  populations than the discs. This correlation is 
  consistent with a picture in which S0 formation comprises the shutting 
  down of star formation in the disc accompanied by a final burst of star 
  formation in the bulge.  Similarly, a trend was found to exist
  whereby galaxies with younger stellar populations have higher 
  metallicities.   The variation in spatial-fit parameters with wavelength 
  also allows us to measure approximate colour 
  gradients in the individual components.  Such gradients were detected 
  separately in both bulges and discs, in the sense that redder light is 
  systematically more centrally concentrated in all components.  However, 
  a search for radial variations in the absorption line strengths determined 
  for the individual components revealed that, although they can be
  sensitively detected where present, they are absent from the vast
  majority of S0 discs and bulges.  The absence of gradients in line
  indices for most galaxies implies that the colour gradient cannot be
  attributed to age or metallicity variations, and is therefore most
  likely associated with varying degrees of obscuration by dust.
  
\end{abstract}

\section{Introduction}\label{sec:introduction}
The evolution of galaxies in different environments is still poorly
understood, as is the relative importance of the different processes
involved. Lenticular galaxies (S0s) lie between spirals and
ellipticals on the Hubble Sequence, and share many properties with
both of these morphologies such as the redder colours and older
stellar populations of ellipticals along with the stellar discs of
spirals. As a result, S0s are often seen as a transitional phase
between spirals and ellipticals, and thus understanding their origins
is thought to be a key stage in understanding galaxy evolution.  The
many influences controlling their evolution are expected to affect the
bulges and discs of these galaxies in different ways, and so the
individual study of these components should provide valuable clues as
to their evolutionary histories.

Evidence for the evolution of galaxies along the Hubble Sequence with
redshift lies in the morphology--density relation for clusters of
galaxies. This relation shows that spirals tend to dominate the
less-dense outer regions, while the early types tend to lie at the
cores of clusters \citep{Dressler_1980}, and that the relative
fraction of spiral galaxies in clusters decreases toward lower
redshifts while the fraction of S0s increases
\citep{Dressler_1997}.  This finding has been supported by
studies using larger samples of galaxies, such as
\citet{Whitmore_1993}, \citet{Fasano_2000} and \citet{Desai_2007}. Similarly,
\citet{Varela_2004} have also found that late-type spirals are more
frequent among isolated galaxies, while S0s are more commonly found in
denser environments such as galaxy groups and clusters.  Many
scenarios have been proposed for the transformation of spirals into
S0s, mostly focusing on the truncation of star formation within the
galaxy followed by passive evolution into an S0. Processes include the
removal of the cold disc gas by ram pressure stripping
\citep{Gunn_1972}, the removal of the hot halo gas by a process usually 
called starvation \citep{Larson_1980,Bekki_2002}, tidal stripping by 
galaxy harassment \citep{Moore_1996,Moore_1998,Moore_1999}, and starbursts 
triggered by unequal mass galaxy mergers \citep{Mihos_1994} and galaxy--cluster
interactions \citep{Merritt_1984,Miller_1986,Byrd_1990}.

Since these processes would affect galactic bulges and discs
differently, many studies have tried looking for clues to S0 star
formation histories by photometric bulge--disc decomposition. The
parameters for the bulge and disc can be plotted as the Kormendy
Relation \citep{Kormendy_1977}, where the effective surface brightness
for each component is plotted against its scale length. This generally
shows that the surface brightness falls as the scale length increases
over a large sample of spirals, but \citet{Barway_2009} found that
while S0s in the field show this trend, faint cluster S0s do not,
instead forming a downward scatter where the effective surface
brightness drops significantly at larger scale lengths. This drop is
thought to be evidence of transformations induced by environmental
effects such as minor mergers, ram pressure stripping or
harassment. \citet{Barway_2007} also found that in faint S0s the
bulge effective radius increases with the disc scale length, while in
brighter S0s it decreases with increasing disc scale length. Secular
evolution of spirals into lenticulars predicts the former trend since
the passive fading of the bulge and disc would conserve their scale
lengths, while the latter correlation would be expected from a more
turbulent transition. Therefore this variation suggests that
lenticulars may have evolved in different ways depending on their
environments.

By analysing multi-waveband photometry it is possible to look at the
colour differences and gradients within the bulge and disc. Such data
can in turn provide information on the star formation histories of
these galaxies since higher metallicities and older stellar
populations strengthen the redder light from the galaxy. Negative
colour gradients have been found in the bulges of S0s
\citep{Terndrup_1994,Peletier_1996} and spirals
\citep{Mollenhoff_2004}, showing that the blue light is distributed
throughout the bulge while redder light is more concentrated in the
centre. In addition to this it has been found that the discs of S0s
and spirals are bluer than the bulges
\citep{Bothun_1990,Peletier_1996,Hudson_2010}, suggesting that disc
galaxies have more recent star formation activity at larger radii
\citep{deJong_1996_2} or higher metallicities in their nuclear regions
\citep{Beckman_1996,Pompei_1997}. These trends indicate the presence
of age and metallicity gradients across the galaxies, but fail to
provide information on whether it represents a gradient within the
individual components, or whether it arises simply from the
superposition of varying amounts of bulge and disc light, where each
component contains stellar populations of different ages and
metallicities.  It also leaves open the possibility that the gradients
could result from differing amounts of dust extinction at different
radii, rather than telling us about spatial variations in the stellar
population.

To resolve this ambiguity, we have developed a new method to apply
bulge--disc decomposition to long-slit spectroscopy of nearby
galaxies. The technique involves dividing the two-dimensional spectrum
into two one-dimensional spectra, one representing purely the bulge
light and the other the disc light.  This decomposition is achieved by
fitting the spatial profile of light at each wavelength with a
disc-plus-bulge model in order to ascertain the fraction of the light
that should be ascribed to each component at that wavelength.  By
repeating this analysis at every wavelength, the individual spectra of
bulge and disc can be constructed.  These ``clean'' spectra can then
be analysed for their ages and metallicities, as well as any gradients
in these quantities, to determine the sequence of events in the
formation of S0 galaxies.

This paper is intended to present the new method, and give an
indication of its potential by applying the technique to a relatively
small sample of S0s from the Fornax Cluster.  Section~\ref{sec:data}
describes the Fornax data sample and
observations. Section~\ref{sec:method} presents the development of the 
method, and its initial application to the Fornax Cluster data.
Section~\ref{sec:ssp} analyses these results in terms of the stellar
populations in the bulges and discs of these galaxies, and
Section~\ref{sec:colourgrad} looks at their colour gradients.
Section~\ref{sec:linegrad} searches for any corresponding gradients in
line indices and hence ages or metallicities.  Finally, our
conclusions are presented in Section~\ref{sec:conclusions}.

\section{The data sample}\label{sec:data}
The initial sample on which we are testing this new method comprises
the nine lenticular galaxies in the Fornax Cluster, as classified by
\citet{Kuntschner_2000}, with inclinations above 60 degrees and
luminosities in the range \mbox{$-22.3<M_{B}<-17.3$}. The data were
obtained using the \mbox{8.2 m} Antu/VLT between 2002 October 2 and
2003 February 24 with the FORS2 instrument in long-slit spectroscopy
mode. The slit was aligned with the major axis of each galaxy, and was
set to 0.5 arcsec wide and 6.8 arcmin in length.  For all galaxies
except NGC~1316, the centre of the galaxy was placed half way along
the slit in order to obtain information along the entire length of the
major axis; NGC~1316 is the largest galaxy in the sample, and it was
necessary to offset it along the slit in order to cover sufficient sky
for background subtraction. The standard resolution collimator was
used in the unbinned readout mode, giving a spatial scale of $0.125\,
{\rm arcsec}$ pixel$^{-1}$, and the GRIS1400V+18 grism gave a
dispersion of 0.318 \AA\ pixel$^{-1}$ over a wavelength range of
\mbox{$4560\le\lambda\le5860$~\AA}. By analysing arc lines, the
spectral resolution was found to be $\sim4$ pixels FWHM, which
corresponds to a velocity resolution of $\sim 73\,{\rm km}\,{\rm
  s}^{-1}$ FWHM or a velocity dispersion of $\sim 31\,{\rm km}\,{\rm
  s}^{-1}$.  Details of the data reduction are given in
\citet{Bedregal_2006a}.

\section{Spectroscopic bulge--disc decomposition}\label{sec:method}

Conventional photometric one-dimensional bulge--disc decomposition of
galaxies involves measuring the luminosity along the galaxy's major
axis. The decomposition is obtained by plotting the luminosity against
radius, and fitting model bulge and disc components to this
profile. The disc is generally modeled as an exponential,
\begin{equation}
I_{D}(R)=I_{D0}\exp(-R/R_{0}),
\label{exponential}
\end{equation} where $I_{D0}$ is the central surface brightness of the
disc and $R_{0}$ is the disc scale length \citep{Freeman_1970}.
Similarly, the light profile of the bulge can be approximated by a de  
Vaucouleurs profile, 
\begin{equation}
I_{B}(R)=I_{Be}\exp\left\{-7.669\left[(R/R_{e})^{1/4} - 1 \right]\right\}, 
\label{deVaucouleurs} 
\end{equation} 
where $R_{e}$ is the bulge effective radius and $I_{Be}$ is the bulge
effective surface brightness \citep{deVaucouleurs_1953}. For a
two-dimensional spectrum we can apply exactly the same method to each
individual wavelength in the data. For example, Fig.~\ref{light
  profiles} shows the light profile along the major axis of NGC~1375
at 5195~\AA, together with the best fit achieved using a de
Vaucouleurs bulge and exponential disc.

 \begin{figure}
   \includegraphics[width=1\linewidth]{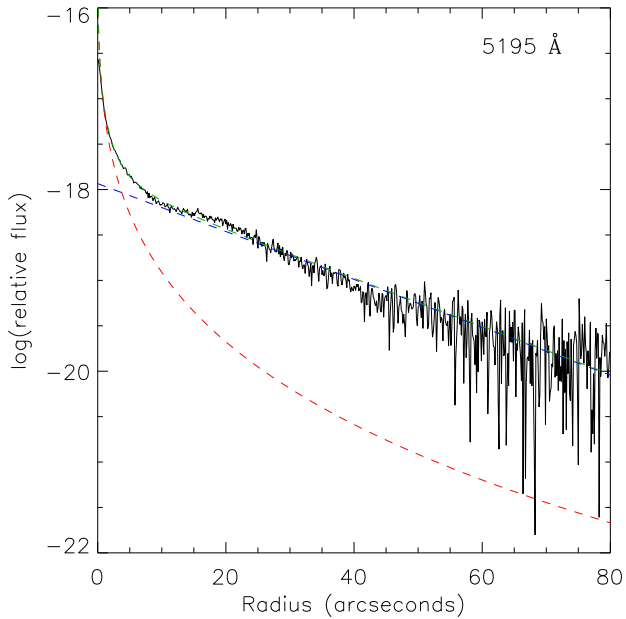}
   \caption{The light profile of NGC~1375 at 5195~\AA\ in black with
     the best fit model in green, comprising the sum of a disc (blue
     line) and bulge (red line) components.\label{light profiles}}
 \end{figure}  

The only complication here is accounting for the velocity dispersion
and radial velocity, where the velocity dispersion of the galaxy
decreases at larger radii, while the rotational velocity red- or
blue-shifts the spectra at larger radii relative to that from the
centre of the galaxy.  Therefore, before fitting the light profile at
each wavelength, the two-dimensional spectrum must be corrected such
that each spectral feature has the same velocity dispersion and radial
velocity, thus ensuring that the light profiles for each wavelength
bin measures the light from the same point on the rest-frame spectrum
at all radii. Extra care must be taken with this correction 
when a bar is present within the galaxy to account for its kinematics.
The velocity dispersion was corrected by convolving the
spectrum from each spatial location with the appropriate Gaussian to
bring it up to the maximum value measured within that galaxy. The
rotational velocity was then corrected by cross correlation, where the
shift in the wavelength of the spectral features at each location was
measured relative to those in the peak spectrum. In order to minimise
the noise in these shift measurements, a rolling average was applied
to the shifts for each spectrum until the radius where the noise
dominated the results, at which point the final reliable shift was
applied to the remaining rows. This gave a smooth velocity curve from
which the necessary shift could be calculated. Since the
two-dimensional spectra cover the entire length of the major axis in
all but one galaxy in the sample, the two halves of each galaxy
spectrum, i.e. the two semi-major axes, were analysed separately.
This duplication was useful to ensure the results were reproducible
for each galaxy, and can also provide information on spatial
asymmetries within the galaxies.

\begin{figure*}
  \includegraphics[width=0.9\linewidth]{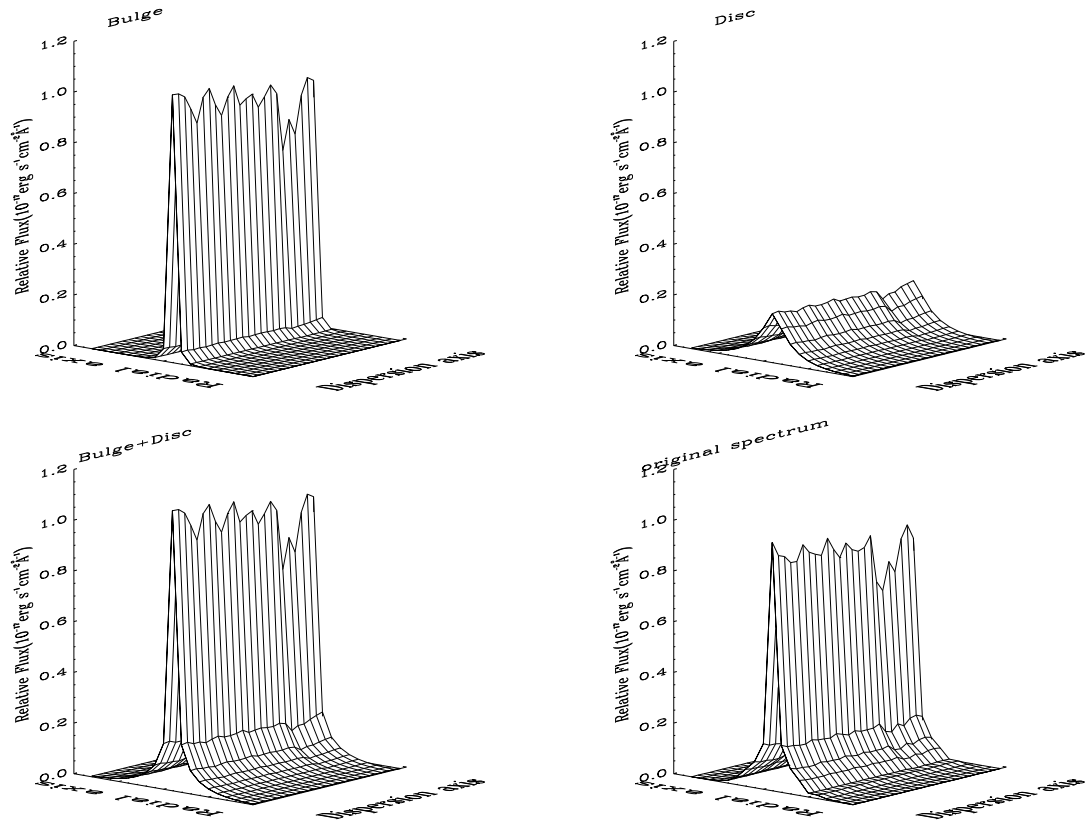}
  \caption{A section of the two-dimensional spectra for the bulge,
    disc and composite models, and the original spectrum of NGC~1381,
    showing how the light profile varies with wavelength. The data
    have been binned such that each step in the radial direction is
    12.5'', and in the dispersion direction is 6.6\AA. The
    wavelength range covered is between 5085\AA\ and 5243\AA, increasing 
    towards the right of the diagrams.  Note that
    the peak values of the bulge and composite spectra have been
    reduced by a factor of four in order to allow them to be plotted
    on the same scale as the original spectrum.\label{surface}}
\end{figure*} 
 
Once this velocity alignment has been carried out, we can perform the
bulge--disc decomposition at all wavelengths.  By way of illustration,
Fig.~\ref{surface} shows a small section of the models produced for NGC~1381
of the bulge, disc and composite galaxy, as derived by fitting the
spatial profile of the two-dimensional spectrum (also shown)
wavelength by wavelength.  In each profile, the central few
arcseconds were masked out prior to fitting in order to eliminate the
effects of seeing, so the peaks of the bulge and disc spectra at very
small radii in Fig.~\ref{surface} were not used in the fit. 
Aditionally, the light profiles of each galaxy were also checked 
by eye to ensure that the exponential parts most likely represented 
the disc rather than the spheroid. This was found to be the case in all 
the galaxies that were decomposed successfully. The results for the bulge 
effective radius and disc scale length for each galaxy from the decomposition 
are given in Appendix~\ref{sec:appendix A}. These were compared to the 
photometric results of \citep{Bedregal_2006a} derived from the Two-Micron All-Sky 
Survey (2MASS) K-band images of these galaxies, and were found to be reasonably 
consistant considering the limited model used in this study.

\begin{figure}
% vskips to try and make LaTeX put figure in sensible place!
\vskip 0.7cm
  \includegraphics[width=1.0\linewidth]{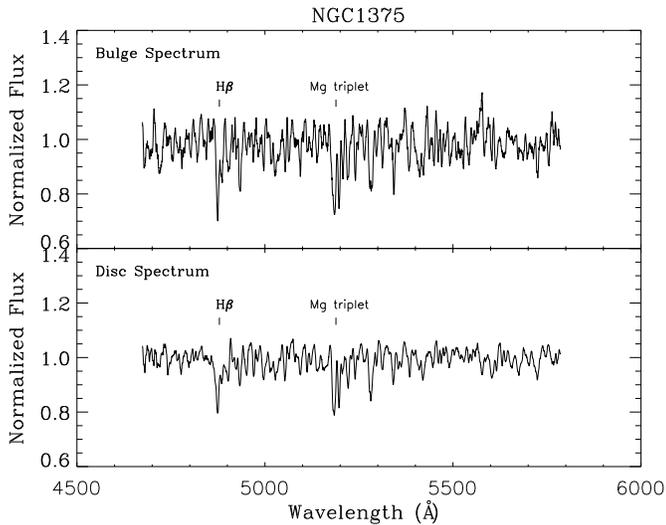}
  \caption{The decomposed one-dimensional bulge and disc spectra for NGC~1375.
    \label{spectra}}
\vskip 0.5cm
\end{figure}

The total luminosity of the bulge and disc at each wavelength can then be
calculated by simple integration, using the standard results that
\begin{equation}
  L_{B}(\lambda)=7.22\pi I_{e}(\lambda)R_{e}^2(\lambda)
\end{equation} 
for the bulge, and 
\begin{equation}
  L_{D}(\lambda)=I_{0}(\lambda)R_{0}^2(\lambda)
\end{equation} 
for the disc, where we have now made the dependence on wavelength,
$\lambda$, of the various fitted parameters explicit.  These
quantities are simply the modeled spectra of the integrated light from
the separate bulge and disc components.  For example,
Fig.~\ref{spectra} shows the spectra derived in this way for NGC~1375.
The high signal-to-noise ratio of the spectra obtained using this
integrated light approach is immediately apparent, such that one can
see that both the H$\beta$ line and the magnesium triplet appear
stronger in the bulge than in the disc; since these features are used
as age and metallicity indicators respectively, this difference
already hints that the bulge contains younger stars with a higher
metallicity than the disc in this galaxy, and we will make a more
quantitative assessment of this impression in Section~\ref{sec:ssp}.

Application of this technique to the full sample of galaxies described
in Section~\ref{sec:data} revealed the kind of data required to implement this
analysis successfully.  In fact, of the nine galaxies, only two
(NGC~1381 and NGC~1375) could be reliably decomposed into bulge and
disc spectra in this manner.  A further three galaxies (IC~1963,
ESO~358-G006 and ESO~359-G002) were found by \citet{Bedregal_2006a} to
have very compact bulges and therefore to be disc dominated from very
small radii, making it impossible to determine a reliable bulge model.
For these systems a disc spectrum was extracted by assuming that the
bulge light was negligible outside of the central masked region, and
just fitting a disc component.  In two further cases (NGC~1380 and
NGC~1316) we found that although the spectra could be decomposed into
bulge and disc components, the resulting model did not reproduce the
original two-dimensional spectrum at all well.  Images of these
galaxies clearly show the presence of major-axis dust lanes in both
cases, which would affect the light profile, and it has also been
suggested by \citet{Caon_1994} that NGC~1316 is a merger remnant,
which would complicate its light profile and therefore make it
unsuitable for fitting a de Vaucouleurs bulge and exponential disc.
The final two spectra (NGC~1380A and ESO~358-G059) could not be
decomposed as their light profiles were again too complex for the
current simple model.  We have therefore ascertained that this
approach works best in systems where both components are well
resolved, and where there is little indication of complicating issues
like strong dust lanes or extra components such as bars.  Fortunately,
comparison between the original two-dimensional spectrum and the model
offers a useful {\it a posteriori} check on the impact of such
additional factors.

\section{Stellar Populations}\label{sec:ssp}

\begin{figure}
  \includegraphics[width=1.0\linewidth]{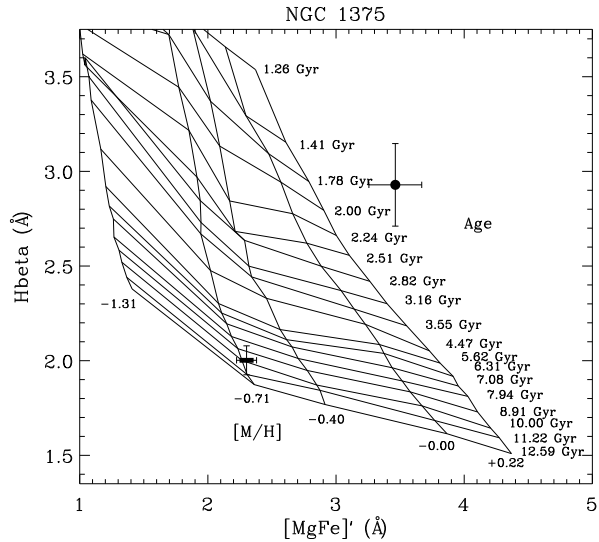}
  \caption{Example of the bulge and disc data for NGC~1375 with the \citet{Vazdekis_2010} SSP model for the bulge kinematics 
	   over-plotted. H$\beta$ is the age indicator while [MgFe]$'$ is the metallicity indicator. The circle represents the 
	   bulge while the rectangle corresponds to the disc value. The error bars represent the statistical uncertainties.
  ~\label{SSP}}
\end{figure}

In the integrated light spectrum of a galaxy, the strength of the
various absorption lines provides information on the underlying
stellar population, with hydrogen lines primarily associated with its
age, while magnesium and iron lines constrain its metallicity.  In
order to obtain quantitative estimates of age and metallicity, these
line indices can be compared to those predicted by simple stellar
population (SSP) models that have been created using stellar libraries
of the same spectral resolution as the data.  In this study, the SSP
models used are those of \citet{Vazdekis_2010}, which uses the
\textsc{miles} stellar library \citep{Sanchez_2006}. 
The library spectra 
have a resolution of $\sim$~58.4~km~s$^{-1}$, and are matched to the 
spectral resolution of the data by convolving the stellar spectra with a 
Gaussian of the appropriate dispersion. These data are available as a web-based
tool\footnote{http://miles.iac.es/} that allows the user to modify the
library spectra to the resolution and redshift of their observations,
and therefore obtain tuned SSP models for their data. 

The line strength indices in the data were measured using the \mbox{\textsc{
indexf}} software of \citet{Cardiel_2007}, which 
uses the Lick/IDS index definitions to calculate a pseudo-continuum over 
each absorption feature based on the level of the spectrum in bands on 
either side \citep{Worthey_1994,Worthey_1997}. The strength of the spectral
feature was then measured relative to this pseudo-continuum, and the 
uncertainty estimated from the propagation of random errors and the 
effect of uncertainties in the radial velocity. 

As one further slight complication, \citet{Bedregal_2008} found that
the galaxies in this sample show traces of emission in the H$\beta$
line which would reduce the absorption-line strength of this line and
skew the results to older ages. To correct for this contamination,
\citet{Gonzalez_1993} identified that the ratio between the equivalent
widths of the [O\textsc{iii}]$_{\lambda5007}$ and H$\beta$ emission
features is around 0.7 in the brightest ellipticals. A later study by
\citet{Trager_2000} found the ratio to vary between 0.33 to 1.25 with
a mean value of 0.6. This mean value was used here to estimate the
level of emission correction necessary for the H$\beta$ index, using
\begin{equation} 
\Delta(\text{H}\beta)=0.6 \times
  \text{EW}[\text{O}\text{\textsc{iii}}]_{\lambda5007},
\end{equation}
where the [O\textsc{iii}]$_{\lambda5007}$ index was measured from the
residual spectrum obtained by subtracting the best combinations of
stellar templates from the original galaxy spectrum. These fits were
achieved using the Penalized Pixel Fitting method (\textsc{ppxf}) of
\citet{Cappellari_2004}, which uses the \textsc{miles} stellar library to model
the line-of-sight velocity distribution as a Gaussian with a series of
Gauss-Hermite polynomials.  For this sample, the H$\beta$ corrections
were never more than 24\% of the H$\beta$ index, with the
majority being less than 10\%, so any residuals from the approximate
nature of this correction are unlikely to compromise the results
significantly. The \textsc{ppxf} code also used the stellar library to 
measure the line of sight velocity and velocity dispersion for each decomposed 
spectrum, which could then be used to tune the SSP models. These values 
are also given in Appendix~\ref{sec:appendix A}.

Having measured and corrected the observed line strengths, SSP models
were created for each decomposed spectrum and plotted using using
H$\beta$ and [MgFe]$'$ as the age and metallicity indicators respectively, 
where the latter was chosen due to its 
negligible dependence on the $\alpha$-element abundance \citep{Gonzalez_1993,
Thomas_2003}.  As an example, Fig.~\ref{SSP} shows the indices derived for 
the bulge and disc spectra of NGC~1375, together with the grid of predictions 
for SSP models of differing ages and metallicities corrected to the 
velocity dispersion of this galaxy, from which one can
read off estimates of the relative age and metallicity for the data.  
Since these measurements were obtained from spectra representing 
the integrated light over the full bulge and disc, they correspond to the global,
luminosity-weighted values for age and metallicity for each component. The errors 
shown reflect the statistical uncertainties outlined 
above. Clearly in this case the bulge is 
inferred to be young and metal rich, while the disc is old and metal poor.

\begin{figure}
  \includegraphics[width=1.0\linewidth]{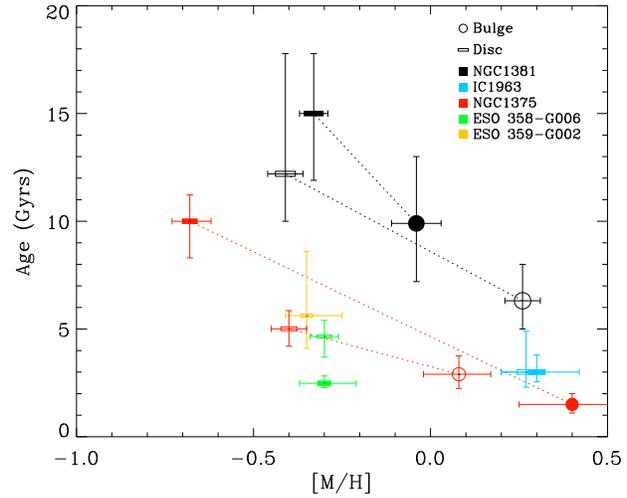}
  \caption{Age versus metallicity for the decomposed spectra. Each
    galaxy is represented by a different colour, with circles and
    rectangles corresponding to bulges and discs respectively.  Dotted
    lines join the bulge and disc of the same galaxy.  The size of
    each symbol reflects the luminosity of the galaxy
    \citep{Madore_1999}.  The filled and open symbols are used to show
    the results from the two semi-major axes in each galaxy, where
    both could be measured.\label{age_z}}
\end{figure}

Figure~\ref{age_z} shows the results of this analysis applied to all
galaxies in the sample.  Where both were observed, we show the two
sides of each galaxy separately, principally as a test of the
reproducibility of the results.  Dotted lines join the bulges and
discs derived from the same side of the same galaxy,  
while points corresponding to different halves of the major 
axis of the same galaxy share the same colour. In the few cases where 
the index measurements lay outside of the SSP model grids, such as in 
Fig.~\ref{SSP}, the metallicities were estimated by using an extrapolation 
method. Due to the uncertainties already mentioned, and the fact that 
different SSP models would give different grids, it is important to 
consider the results in Fig.~\ref{age_z} as constraining the relative ages 
and metallicities within the data set as opposed to 
their absolute values. The errors shown on the data points correspond to 
the statistical errors given in Fig.~\ref{SSP} plus interpolation errors, 
while a more realistic measure of the uncertainties can be obtained by 
comparing the results for each half of the major axis of each galaxy, 
as represented by closed and filled symbols. It can be seen that both 
within individual galaxies and viewing the data set as a whole, there 
is clearly a trend in the sense that the bulges are systematically younger 
and more metal rich than the discs.

\section{Colour Gradients}\label{sec:colourgrad}

\begin{figure}
  \includegraphics[width=1.0\linewidth]{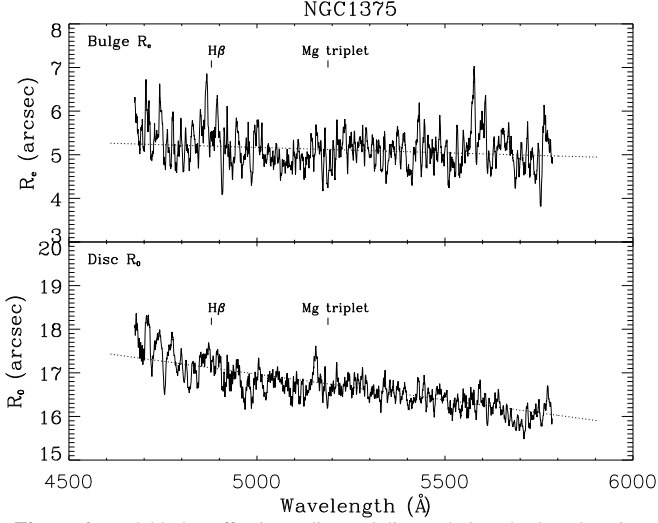}
  \caption{\small{Model bulge effective radius and disc scale length
      plotted against wavelength for NGC1375.  A strong negative
      gradient can be seen in the disc scale-length, corresponding to
      a negative colour gradient across the disc, while the gradient
      in the bulge is much weaker.}  ~\label{radii}}
\end{figure}

As well as estimating the global ages and metallicities of the bulges
and discs, this spectral decomposition method also provides an 
approximate measure of any gradients in the properties of components.  
Specifically, if there were a colour gradient in a component, such that, 
say, red light was more centrally concentrated than blue light, then the
characteristic size-scale of the component would be smaller in red
light than in blue light.  Because we determine the length-scales of
both disc and bulge as a function of wavelength, it is straightforward
to see whether any such gradients exist.  For example,
Fig.~\ref{radii} shows the bulge and disc length-scales across the
range of wavelengths observed for NGC~1375.  In both cases, there is a
negative gradient, indicating that the centre of each component is
redder than its outer parts, although the effect is clearly stronger in
the disc than the bulge.  

In order to quantify this effect in terms of more conventional colour
gradients, we can calculate the bulge effective radius and the disc
scale-length at the the central wavelengths of the $B$ and $V$ band
filters from the Johnson-Cousins system [\mbox{4450 \AA} and
\mbox{5510 \AA} respectively \citep{Bessell_1990}]. For the bulge, the
ratio of light in the $B$-band to that in the $V$-band is given by
\begin{equation}
  \frac{I_{B}}{I_{V}}=\bigg(\frac{I_{eB}}{I_{eV}}\bigg)\exp\Bigg\{-7.67\bigg[\bigg(\frac{1}{R_{eB}}\bigg)^{
      1/4}-\bigg(\frac{1}{R_{eV}}\bigg)^{1/4}\bigg]R^{1/4}\Bigg\},
\label{BVdeVaucouleurs}
\end{equation}
or, in magnitudes,
\begin{align} 
 B-V & = {\rm const} - 2.5\log_{10}{\bigg(\frac{I_{eB}}{I_{eV}}\bigg)}\nonumber \\
     & + 19.18\Bigg(\frac{1}{R_{eB}^{1/4}}-\frac{1}{R_{eV}^{1/4}}\Bigg)R^{1/4}\log_{10}{e}. 
\label{magdeVaucouleurs}
\end{align}
Differentiating with respect to $R^{1/4}$, we find 
\begin{equation} 
\frac{d(B-V)}{d(R^{1/4})}=8.3\Bigg(\frac{1}{R_{eB}^{1/4}}-\frac{1}{R_{eV}^{1/4}}\Bigg).
\label{BVdifferential}
\end{equation}
This equation can then be integrated out from the centre to any
radius, to obtain a change in colour, 
\begin{equation} 
(B-V)_{R}-(B-V)_{0}=8.3\Bigg(\frac{1}{R_{eB}^{1/4}}-\frac{1}{R_{eV}^{1/4}}\Bigg)R^{1/4}.
\label{BVintegral}
\end{equation}
As a characteristic radius, we set $R=R_{eV}$ in
Equation~\eqref{BVintegral}, and then divide by $\log R_{eV}$ to create an
appropriate gradient quantity.  

We can also obtain a set of analogous equations in order to define a
gradient in the disc:
\begin{equation} 
\frac{I_{B}}{I_{V}}
 = \bigg(\frac{I_{0B}}{I_{0V}}\bigg)\exp\Bigg[-\bigg(\frac{1}{R_{0B}}-\frac{1}{R_{0V}}\bigg)R\Bigg];
\label{BVexponential}
\end{equation}
\begin{align} 
B-V & ={\rm const}-2.5\log_{10}{\bigg(\frac{I_{0B}}{I_{0V}}\bigg)}\nonumber \\
    & +2.5\Bigg(\frac{1}{R_{0B}}-\frac{1}{R_{0V}}\Bigg)R\log_{10}{e}; 
\label{magexponential}
\end{align}
\begin{equation} 
\frac{d(B-V)}{d(R)}=1.09\Bigg(\frac{1}{R_{0B}}-\frac{1}{R_{0V}}\Bigg);
\label{BVdifferential2}
\end{equation}
\begin{equation} 
(B-V)_{R}-(B-V)_{0}=1.09\Bigg(\frac{1}{R_{0B}}-\frac{1}{R_{0V}}\Bigg)R.
\label{BVintegral2}
\end{equation}

\begin{figure}
  \includegraphics[width=1.0\linewidth]{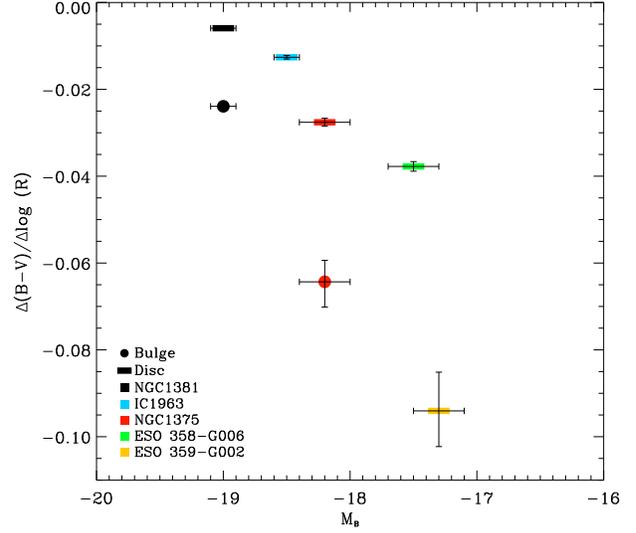}
  \caption{The colour gradients in the bulges (circles) and discs (rectangles) of each galaxy against the galaxy's absolute $B$-band 
   magnitude. The colours of the points are as in Fig.~\ref{age_z}.
  \label{colour_grad}}
\end{figure}

\begin{figure*}									%moved here to keep in correct place in dvi file
  \includegraphics[width=1.0\linewidth]{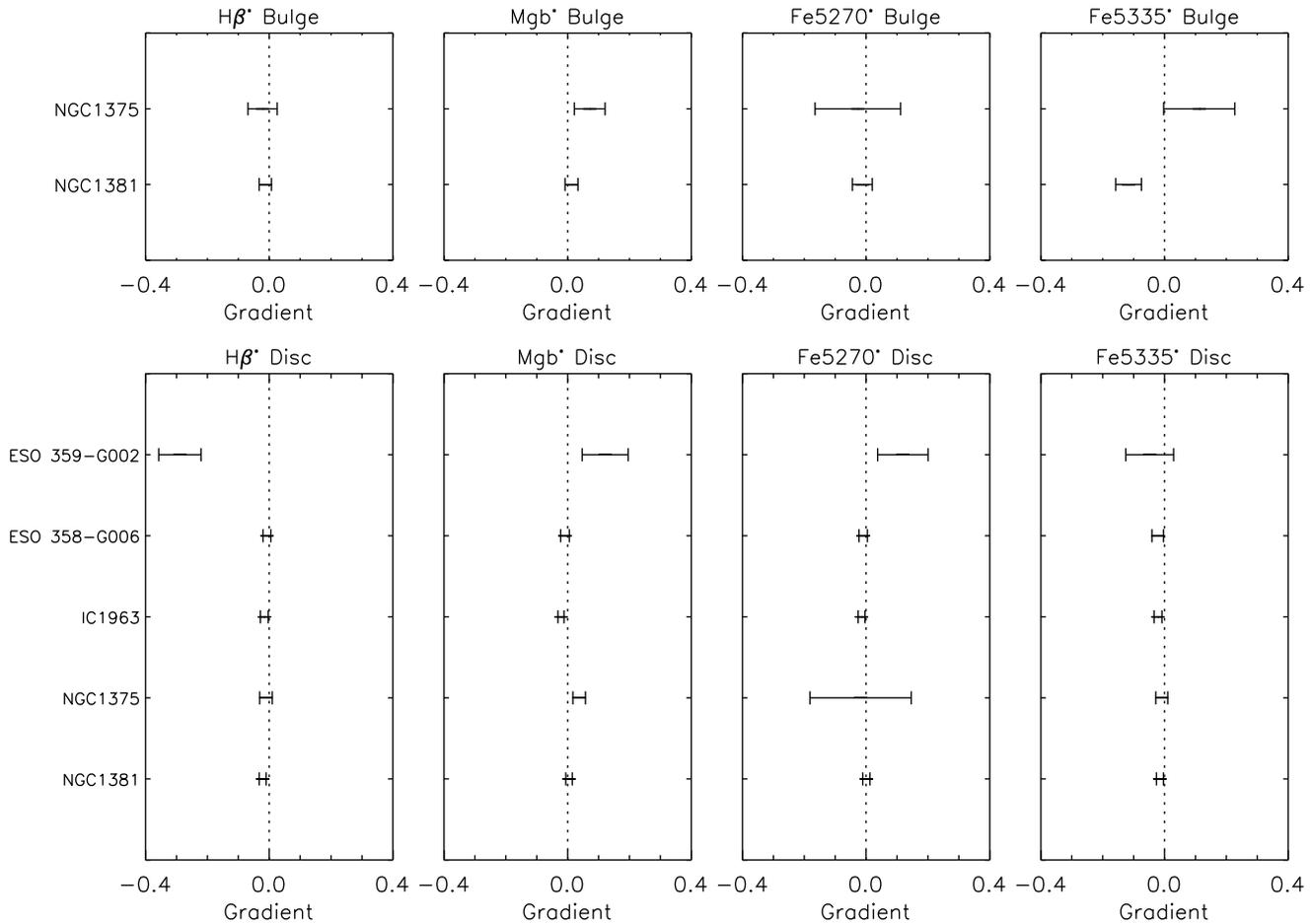}
  \caption{The line gradients for the bulge and disc in each galaxy,
    calculated using $\Delta \text{Index}^{*}/\Delta\log_{10}(R)$.}
    \label{index_gradients}
\end{figure*}

The colour gradients obtained from these formulae using the observed
variation in scale-length with wavelength for each component of each
galaxy are shown in Fig.~\ref{colour_grad}.  In all cases, the colour
gradients are negative, indicating that the centres are redder than
the outskirts, in both bulge and disc components.  There also appears
to be a trend that the gradients are stronger in fainter galaxies.
The range of values obtained are directly comparable to those found in
previous studies of early-type galaxies, with \citet{LaBarbara_2009}
reporting a typical gradient of \mbox{$\frac{\Delta g-r}{\Delta \log
    R}\sim-0.071\pm0.003$}, \citet{Roche_2010} finding a range of
\mbox{$-0.8<\frac{ \Delta g-r}{\Delta \log R}<0.0$}, and
\citet{Suh_2010} reporting values of \mbox{$-0.4<\frac{\Delta
    g-r}{\Delta \log R}<0.0$}.  \citet{denBrok_2011} similarly
detected gradients within the range \mbox{$-0.2<\frac{\Delta
    g-i}{\Delta \log R}<0.0$} for a sample of early-type galaxies, but
found that S0s show weaker gradients than ellipticals, suggesting that
the underlying mechanisms may differ.

Attempts have also been made to ascertain the causes of such
gradients.  For example, \citet{LaBarbara_2009} concluded that the
main contributors to the colour gradients were the radial variations
in metallicity over the galaxies, and that, while a small positive age
gradient was also present, its contribution to the colour gradient was
negligible in comparison.  The spectral decomposition method presented
here provides us with a new mechanism for seeing directly in the
individual components whether these gradients are associated with
changes in age or metallicity, and we now describe how this
information can be exploited.

\section{Line Index Gradients}\label{sec:linegrad}

If there were a gradient in the strength of a particular
absorption line within a single component, one would expect that the
scale-length as determined within the absorption feature would differ
from that of the surrounding continuum, which, in turn, would show up
in the spectrum of that component generated from the best-fit model,
like those shown in Fig.~\ref{surface}.  We have therefore analysed
these model spectra by calculating the Lick indices from them as a
function of radius.  A logarithmic gradient was then generated for
each index by calculating the magnitude version of the indices, 
\begin{equation} 
\text{Index}^{*}=-2.5\log_{10}\Bigg(1-\frac{\text{Index}}{\Delta\lambda}\Bigg),
\end{equation} 
and measuring its variation with $\Delta\log_{10}(R)$. 
The resulting gradients for the bulge and disc
of each galaxy are presented in Fig.~\ref{index_gradients}.

Since any signal in this plot arises from a variation in the
component's scale-length at the wavelength of the index, the error on
each measurement was assessed by calculating the indices that one
obtains by treating the plots of scale-length versus wavelength (such
as those shown in Fig.~\ref{radii}) as spectra, and extracting Lick
indices from them: if there were a signal, then the value of
scale-length in the central band should differ significantly from the
value in the pseudo-continuum, leading to a non-zero index.  To measure
the significance of any such measurement, and hence the appropriate
size of the error bars in Fig.~\ref{index_gradients}, the indices
measured at the true index wavelengths in plots like Fig.~\ref{radii}
were compared to those obtained from random locations in these plots. 
 These results were found to be consistant with those of 
\citet{Bedregal_2011}, in which the line strength gradients
were measured over the whole galaxy. 

\begin{figure}		
  \includegraphics[width=1.0\linewidth]{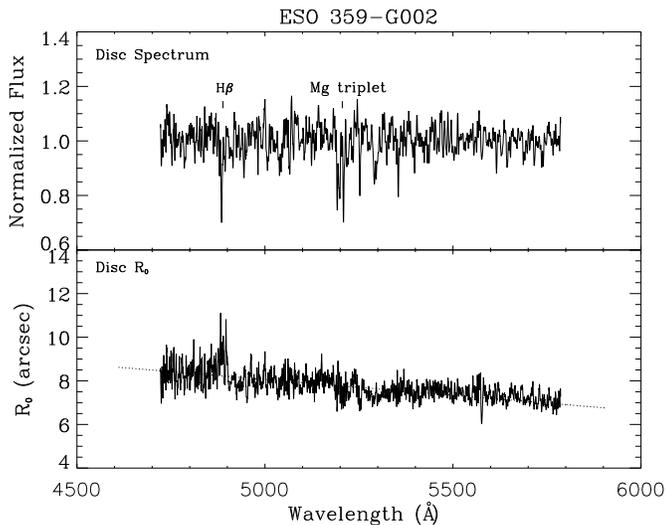}
  \caption{The decomposed disc spectrum for ESO~359-G002 (top) and  
  a plot of its disc scale length against wavelength (bottom). A feature 
  is present at the wavelength of the H$\beta$ line on the scale-length plot,
  implying that there is a gradient in its line strength.
  \label{ESO002_R0feature}}
\end{figure}

\begin{figure}		
  \includegraphics[width=1.0\linewidth]{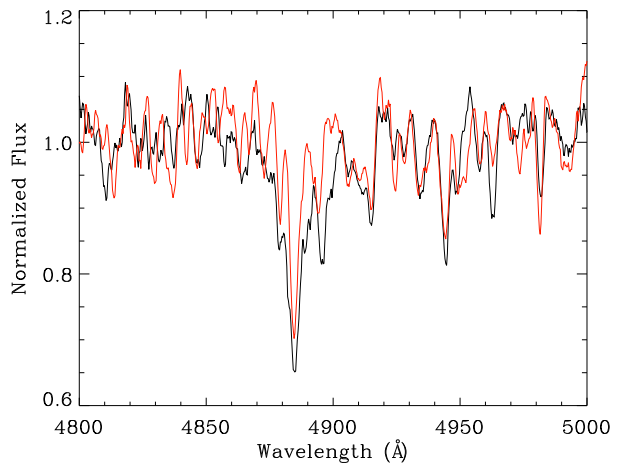}
  \caption{ The H$\beta$ feature in the reconstructed spectra
    of the disc of ESO~359-G002 at the radii of the inner seeing limit
    (black), and the disc scale length (red).}
    \label{ESO002_Hb}
\end{figure}

As can be seen from Fig.~\ref{index_gradients}, the H$\beta$ index of
ESO~359-G002 shows a gradient detected with a $> 4\sigma$
significance, which would suggest the presence of an age variation
across its disc.  The origins of this detection can be seen in
Fig.~\ref{ESO002_R0feature}, where the scale length measured at the
wavelength of H$\beta$ differs from that in the surrounding continuum,
just as described above.  The effect of this variation on the
reconstructed spectra is illustrated in Fig.~\ref{ESO002_Hb}, which
shows the derived H$\beta$ feature at radii of $2\,{\rm arcsec}$ (the
inner seeing limit of the galaxy) and at $8\,{\rm arcsec}$ (the disc
scale length).  The SSP model for this galaxy was used to translate
the H$\beta$ line index gradient into an approximate age gradient of
$\Delta\log_{10}({\rm age})/\Delta\log_{10}(R) \sim 1.1 \pm 0.2$. This
corresponds to ages of $1.86 \pm0.14$~Gyrs and $8.77 \pm0.16$~Gyrs at
the respective inner and outer radii shown in Fig.~\ref{ESO002_Hb}.
Note, however, that such an age gradient is in the wrong sense to
explain the colour gradient seen in this system.

In trying to explain the observed colour gradients, the situation 
is no more helpful in all the other galaxies.  As is evident from 
Fig.~\ref{index_gradients}, none of the components in any
of these systems display significant gradients in any of their
indices.  We can translate these limits into limits on the variation
in the properties of the stellar population between the centre and
each component's characteristic radius ($R_e$ for bulges and $R_0$ for
discs) as above.  Typically, this analysis yields an upper limit of a
$\sim 30\%$ change in age and metallicity between these locations.
This corresponds to a colour gradient of \mbox{$\big\lvert
\frac{ \Delta (B-V)}{\Delta \log R}\big\rvert\approx 0.02$} using 
the models of \citet{Bruzual_2003}, which cannot
explain the sizable colour gradients found in Section~\ref{sec:colourgrad}.  
It would therefore appear that some other factor, most likely dust 
reddening that varies with radius, must be the underlying cause of 
the colour gradients.


\section{Discussion}\label{sec:conclusions}

We have presented a new method for analysing S0 galaxy spectra by
decomposing their major-axis light into bulge and disc components on a
wavelength-by-wavelength basis, in order to construct clean, high-quality
spectra for each individual component.  Application of this method to
a preliminary sample of data from the Fornax Cluster has revealed that
as long as the galaxies are well-described by this two component
model, and the bulge region is sufficiently resolved to allow a
reliable structural fit, it is quite possible to spectrally decompose
galaxies in this way.  

Even this initial small sample reveals some clear systematic trends in
the properties of Fornax S0s.  Specifically, analysis of the Lick
indices of the individual components shows that in the two galaxies 
that could be decomposed into two components, the bulges are
systematically younger and more metal rich than the discs.  This is 
also found to be the case more globally when comparing galaxies across the 
sample. Since these quantities are luminosity-weighted averages for the 
individual components, we cannot rule out the probability that the differences 
arise from relatively small amounts of late star formation, but they nonetheless 
point to real diffeences in star formation history. The positive age gradient
between the bulge region and disc region has previously been
documented in this data set by \citet{Bedregal_2011}, and has been
seen in other lenticular galaxies by \citet{Fisher_1996},
\citet{Bell_2000}, \citet{Kuntschner_2000}, \citet{MacArthur_2004} and
\citet{Prochaska_2011}, but the new analysis technique allows us for
the first time to cleanly measure the relative ages of the 
individual components, and also to optimise the signal-to-noise ratio of 
the data, and hence the precision of the measurements, by combining all 
of the two-dimensional spectral data into just two one-dimensional
spectra.

By analysing the variation in the bulge and disc components'
scale-lengths with wavelength, we were also able to make 
approximate measurements of the colour gradients in the individual
components, which turned out to be systematically negative, with the
centre of each component redder than its outskirts.  Again, such
gradients have been seen before, but here for the first time we can
show clearly that they exist in the individual components, not just
arising from the superposition of monochromatic individual components
whose relative contributions vary with radius.  

By analysing any variations in the best-fit structural parameters at
the wavelengths of the various Lick indices, we have further been able
to measure gradients in age and metallicity. In only one galaxy was a
significant age gradient detected, which interestingly was found to be
in the wrong sense to explain the system's colour gradient.  In all
other components of all other galaxies, no discernable gradients were
detected.  The upper limits these measurements imposed were also
sufficient to rule out the possibility that any of the observed colour
gradients could be due to variations in the stellar population.  It
therefore seems likely that the red centres of all components must be
attributed to centrally-concentrated dust in these systems.

Although the first results presented here already illustrate the
potential power of spectroscopic bulge--disc decomposition, there are
still plenty of areas for further development.  One could extend the
model by, for example, introducing a more general S\'ersic profile
instead of the de Vaucouleurs law to fit a wider range of bulge types,
or one could introduce to the model additional components such as a
bar, in order to see where the stellar population properties of such
components fit into the developing picture of S0 galaxy formation.

It would also be helpful to apply this analysis to a somewhat larger
sample obtained in a different cluster, both to improve the statistics
and to search for variations between S0 properties in different
locations.  To this end, we are currently analysing long-slit spectra
from a further sample of 21 S0 galaxies in the Virgo Cluster, and look
forward to presenting the results shortly.

In addition, the method could be straight-forwardly extended 
to IFU data, where the effects of small scale variations, such as those 
due to dust lanes, can be reduced, and the star formation history 
throughout the two-dimensional structure can be measured more robustly. 
We are currently modifying this technique in order to apply it to such data.

\section*{Acknowledgements}
We thank the anonymous referee for the careful reading of the 
manuscript, and the range of helpful suggestions. We would also like to 
thank Omar Almaini for his useful input to this study, and Patricia 
S\'anchez-Bl\'azquez for very helpful discussions and for independently 
testing our initial results. We would also like to thank Michele Cappellari 
for the use of his \textsc{ppxf} software. This work was based on 
observations made with ESO telescopes at Paranal Observatory under the 
program ID 070.A-0332, and was supported by STFC.

\bibliographystyle{mn2e}

\bibliography{myrefs}

\appendix
\section{Tables}\label{sec:appendix A}
In this appendix we include tables with different parameters for each 
galaxy in the sample. Only NGC~1381 and NGC~1375 could be decomposed into bulge 
and disc components, IC~1963, ESO~358-G006 and ESO~359-G002 gave disc only fits, 
and NGC~1316, NGC~1380, NGC~1380A and ESO~358-G059 could not be fitted with the 
current model. The errors are presented in parentheses next to the values, 
and represent one sigma uncertainties in the measurements. The results 
for $R_{e}$, $R_{0}$, $V_{LOS}$ and $\sigma$ are given for both halves of the 
major axis in each case, where $V_{LOS}$ and $\sigma$ are the line-of-sight 
velocity and velocity dispersions after the corrections described in 
Section~\ref{sec:method}.\\
\\

% \newpage

{\renewcommand{\arraystretch}{1.5}
\renewcommand{\tabcolsep}{0.57cm}
\begin{tabular*}{0.9\textwidth}{ l c c c c c}
\hline \hline
 Name & $M_{B}$  & $R_{e}$ & $R_{0}$ & $\sigma_{0}$ & $V_{LOS}$\\
      &          & [arcsec] & [arcsec] & [km s$^{-1}$] & [km s$^{-1}$] \\
 (1)  &    (2)   & (3) & (4) & (5) & (6) \\
\hline \hline
NGC1381 & $-19.1$ (0.1) & 6.7 (0.2) & 25.1 (0.2) & 150.5 (12.6) & 1718.5 (8.8)\\
        &               & 7.5 (0.2) & 26.4 (0.3) & 157.9 (14.0) & 1715.0 (9.0)\\
\hline
NGC1375 & $-18.2$ (0.2) & 5.0 (0.4) & 16.4 (0.3) & 115.0 (9.9) & 785.2 (6.6)\\
        &               & 3.5 (0.4) & 16.8 (0.5) & 116.5 (9.4) & 778.3 (6.5)\\
\hline
IC1963 & $-18.5$ (0.1) & -- & 15.5 (0.3)  & 59.0 (2.3) & 1635.5 (1.7)\\			%edit from 43.5
       &               & -- &  15.0 (0.3) & 61.8(1.8) & 1631.7 (1.6)\\			%edit from 77.2
\hline
ESO 358-G006 & $-17.5$ (0.2) & -- & 10.5 (0.2) & 98.8 (5.7) & 1264.2 (4.3)\\
             &               & -- & 10.4 (0.2) & 99.6 (6.4) & 1258.5 (4.7)\\
\hline
ESO 359-G002 & $-17.3$ (0.2) & -- & 7.3 (0.5) & 58.9 (3.3) & 1439.7 (2.5)\\
             &               & -- & 7.4 (0.5) & 62.7 (3.2) & 1457.3 (2.6)\\
\hline
NGC 1316 & $-22.3$ (0.1) & -- & -- & -- & -- \\
\hline
NGC 1380 & $-20.6$ (0.1) & -- & -- & -- & -- \\
\hline
NGC 1380A & $-18.1$ (0.2) & -- & -- & -- & -- \\
\hline
ESO 358-G059 & $-17.4$ (0.2) & -- & -- & -- & -- \\

\hline \hline
\multicolumn{6}{p{6in}}{
Note. Column (2): absolute magnitude in B-band according to \citet{Madore_1999};
Column (3): bulge effective radius, measured at the central wavelength of the 
V-band filter (5510~\AA); Column (4): disc scale length, measured at the central 
wavelength of the V-band filter (5510~\AA); Column (5): velocity dispersion; 
Column (6): line-of-sight velocity.
}
\end{tabular*}}

\end{document}